\documentclass{article}

\usepackage{epsf}

\def\ps{\noindent}

\def\apj{Astrophys. J.}

\def\apjs{Astrophys. J. Suppl.}
\def\mnras{M.N.R.A.S.}

\def\vc#1{\mathbf{#1}}
\def\Mpc{~{\rm Mpc}}
\def\Mpch{h^{-1}{\rm Mpc}}

\def\etal{et al.}

\begin{document}

\begin{center}
\begin{LARGE}
Particle-Mesh code for cosmological simulations
\end{LARGE}

\textbf{Anatoly Klypin and Jon Holtzman}

{Department of Astronomy, New Mexico State University}

{Las Cruces, NM 88001}
\end{center}

\centerline{\textbf{Abstract}}

Particle-Mesh (PM) codes are still very useful tools for testing
predictions of cosmological models in cases when extra high resolution
is not very important. We release for public use a cosmological PM
N-body code. The code is very fast and simple.
We provide a complete package of routines needed to set initial
conditions, to run the code, and to analyze the results.  The package
allows you to simulate models with numerous combinations of parameters:
open/flat/closed background, with or without the cosmological constant,
different values of the Hubble constant, with or without hot neutrinos,
tilted or non-tilted initial spectra, different amount of baryons. 

Routines are included to measure the power spectrum and the density
distribution function in your simulations, and a bound-density-maxima
code for halo finding. We also provide results of test runs. A
simulation with $256^3$ mesh and $128^3$ particles can be done in a
couple of days on a typical workstation (about 70Mb of memory is
needed). To run simulations with $800^3$ mesh and $256^3$ particles one
needs a computer with 1Gb memory and 1Gb disk space. The code has been
successfully tested on an HP workstation and on a Sun workstation running
Solaris, but we expect it should work on other systems.

The package can be downloaded from 
{http://astro.nmsu.edu/$\sim$aklypin/PM/PMcode.tar.gz}
A PostScript version of {this manual} can be obtained from 
{http://astro.nmsu.edu/$\sim$aklypin/PM/PMcode.ps.gz}

We provide this tool as a service to the astronomical community, but we
cannot guarantee results.

\section{\textbf{Introduction}}

There are many different numerical techniques to follow the evolution
of a system of many particles. For earlier reviews see Hockney \&
Eastwood (1981) and Sellwood (1987). The most frequently used methods
for cosmological applications fall in three classes: Particle Mesh (PM)
codes, Particle-Particle/Particle-Mesh (P$^3$M) codes, and TREE codes.
All methods have their advantages and disadvantages.

\textbf{PM codes} use a mesh for the density and potential.  As a
result, their resolution is limited by the size of the mesh. The largest
simulations done by the author have been on a $800^3$ mesh with $3\times 256^3
=1.5\times 10^8$ particles. The SP2 parallel supercomputer at Cornell has
been used to run simulations with a $1600^3=4.096\times 10^9$ mesh 
(Gross, 1997).
There are two advantages of the method: i) it is fast (the smallest
number of operations per particle per time step of all the other
methods), ii) it typically uses very large number of particles. The
latter can be crucial for some applications. There are a few variants of
PM codes.  A ``plain-vanilla'' PM was described by Hockney \& Eastwood
(1981), and this includes a Cloud-In-Cell density assignment and a 7-point
discrete analog of the laplacian operator. Higher order approximations
improve the accuracy on large distances, but degrade the resolution
(e.g. Gelb (1992)). In an effort to reduce the order of approximation
and to increase the resolution, Melott (1986) introduced the staggered
mesh. It gives a better resolution on cell-size distances, but
particles get self-forces (an isolated particle experiences a force
from itself), which might be not a welcome feature.

\textbf{P$^3$M codes} are described in detail in Hockney \& Eastwood
(1981) and in Efstathiou \etal (1985). They have two parts: a PM part,
which takes care of large-scale forces, and a PP part, which adds a
small-scale particle-particle contribution. The simulations usually
have $64^3$--$100^3$ particles.  Because of strong clustering at late
stages of evolution, the PP part becomes prohibitively expensive once
large objects start to form in large numbers. Significant speed is
achieved in a modified version of the code which introduces subgrids
(next levels of PM) in areas with high density (AP$^3$M code of
Couchman (1991)). With modification this code runs as fast as TREE code
even for heavily clustered configurations (Couchman 1991).

\textbf{TREE codes} are the most flexible code in the sense of the choice of
boundary conditions (Appel 1985, Barnes \& Hut 1986, Hernquist
1987). They are also more expensive than PM: it takes 10-50 times more
operations.  Bouchet \& Hernquist (1986) and Hernquist, Bouchet \& Suto
(1991) have extended the code for the periodical boundary conditions, which
are important for simulating large-scale fluctuations.

\textbf{Multigrid methods} were introduced long ago, but only recently
have they started to show a potential to produce real results (Anninos,
Norman \& Clarke 1994, Suisalu  \& Saar 1995, Kravtsov \etal 1997).  At
present the most advanced and fastest multigrid code has been developed
by Kravtsov \etal (1997).

\section{\textbf{ Equations and dimensionless variables}}

The equations of motion of particles in expanding
coordinates, which are used in our PM code, were presented by Kates
\etal (1991). Different numerical effects (including resolution) were
discussed in  Klypin \etal (1996).  We use comoving
coordinates ofr particles $\vc{x}=\vc{x}(t)$, which are related to
proper coordinates by $\vc{r}=a(t)\vc{x}$, where $a(t)=(1+z)^{-1}$ is
the expansion parameter.  Instead of using peculiar velocity
$\vc{v}_{\rm pec}=a\dot\vc{x}$ we write the equations for particle momenta
$\vc{p}$: 

\begin{equation}
\vc{p}=a^2\dot\mathbf{x}, \quad \vc{v}_{\rm pec}=\vc{p}/a  \label{Eqp}
\end{equation}

 This choice of ``velocity''
simplifies the equations of motion by removing a few terms with $\dot a/a$. It
is also convenient to change the time variable from time $t$ to the
expansion parameter $a$. The equations governing the motion of particles
are:
\begin{equation}
{d\vc{p}\over da} =-{\nabla\phi\over\dot a}, \quad
{d\vc{x}\over da} = {\vc{p}\over \dot a a^2}         \label{Eqdxda}
\end{equation}

\begin{equation}
\nabla^2\phi = 4\pi G\Omega_{\rm m}(t)a^2\rho_{\rm cr}(t)\delta
=4\pi G\Omega_0\rho_{\rm cr,0}{\delta\over a}, \quad
\delta\equiv {\rho(\vc{x})-\rho_b\over\rho_b},       \label{EqPoisson}
\end{equation}

\begin{equation}
\dot a\sqrt{a} =H_0\sqrt{\Omega_0 
         +\Omega_{\rm curv,0}a
         +\Omega_{\Lambda,0}a^3},
	\quad \Omega_0+\Omega_{\rm curv,0}+ \Omega_{\Lambda,0} =1, \label{Eqpart}
\end{equation}

\noindent
where $\Omega_0=\Omega_m(z=0)$, $\Omega_{\rm curv,0}$, and
$\Omega_{\Lambda,0}$ are the density of the matter, effective
density of the curvature, and the cosmological constant in units of the
critical density at $z=0$. The curvature contribution is positive for
negative curvature.

Dimensionless variables (shown with tildas below) are defined by
introducing the length of a cell of the grid $x_0$ and by measuring
the time in units of $1/H_0$:

\begin{equation}
\vc{x} = x_0\tilde\vc{x}, \quad t=\tilde t/H_0,   \label{Eqtime}
\end{equation}

\begin{equation}
\vc{v}_{\rm pec}=(x_0H_0)\tilde\vc{p}/a, 
	\quad\phi=\tilde\phi(x_0H_0)^2            \label{Eqvel}
\end{equation}

\begin{equation}
\rho={\tilde\rho\over a^3}{3H_0^2\over 8\pi G}\Omega_0  \label{Eqrho}
\end{equation}

Equations (2--5) can be rewritten in terms of dimensionless variables:
\begin{equation}
{d\tilde\vc{p}\over da}= -F(a)\tilde\nabla \tilde\phi,
\quad {d\tilde\vc{x}\over da}= F(a){\tilde\vc{p}\over a^2}  \label{Eqtildep}
\end{equation}

\begin{equation}
\tilde\nabla^2 \tilde\phi ={3\over 2} {\Omega_0\over a}
 (\tilde\rho -1),        \label{Eqtildephi}
\end{equation}
where
\begin{equation}
        F(a)\equiv H_0/\dot a =\left({ 
	\Omega_0 +\Omega_{\rm curv,0}a+\Omega_{\Lambda,0}a^3}
	\over a \right)^{-1/2}.   \label{EqFa}
\end{equation}

\noindent
Equations (\ref{Eqtildep} -- \ref{Eqtildephi}) are solved numerically by
the PM code.

If $L$ is the length of the computational box at $z=0$, $N_{\rm grid}$ is the
number of grid cells in one direction, and $N_{\rm row}$ is the number
of particles in one direction, which contribute a fraction $\Omega_0$ of
the critical density, then the transformations from dimensionless variables
given by the code to dimensional variables are given by
\begin{equation}
x=x_0\tilde x, \quad
   x_0 ={L\over N_{\rm grid}} 
       =7.8{\rm kpc}\left({L_{\rm Mpc}\over N_{\rm grid}/128}\right),
\label{Eqxnot} 
\end{equation}

\begin{equation}
   v_{\rm pec} = (x_0H_0){\tilde p\over a} = 0.781{\rm km\over s}\cdot
	{\tilde p\over a}\cdot {L_{\rm Mpc}h\over  N_{\rm grid}/128},
\label{Eqvscale}
\end{equation}

\begin{equation}
{\rm Mass} =N_{\rm particles}\cdot m_1,
 \quad m_1 =\Omega_0\rho_{\rm cr,0}\left({L\over N_{\rm row}}\right)^3
           =1.32\cdot 10^5(\Omega_0h^2)
\left({L_{\rm Mpc}\over N_{\rm row}/128}\right)^3
\label{Eqmass}
\end{equation}

\section{\textbf{Scheme of integration}}

Equations (\ref{Eqtildep} -- \ref{Eqtildephi}) are solved using finite
differences with a constant step in space $\Delta x =\Delta y =\Delta
z =1$ and a constant step in the expansion parameter $\Delta a$. We
use the ``leap-frog'' scheme to advance coordinates and velocities
from one moment to another. (In the following we drop tildas for all
dimensionless variables.) At any moment $a_n=a_0+n\Delta a$, we have the
coordinates $\vc{x}_n$ and the potential $\phi_n$. Velocities
$\vc{p}_{n-1/2}$ are defined at $a_{n-1/2}=a_n-\Delta
a/2$. The coordinates and the velocities for the next moment are found using:

\begin{eqnarray}
\vc{p}_{n+1/2} &=&\vc{p}_{n-1/2}-F(a_n)\nabla\phi_n\Delta a,\nonumber\\
\vc{x}_{n+1} &=&\vc{x}_n+{F(a_{n+1/2})\over a_{n+1/2}^2}
\vc{p}_{n+1/2}\Delta a
\label{Eqmove}
\end{eqnarray}

\noindent
In order to solve eq.(\ref{Eqtildephi}) we approximate the Laplacian operator
using the 7-point ``crest'' template: 
\begin{equation}
\nabla^2\phi\approx
\phi_{i\pm 1,j,k}+\phi_{i,j\pm 1,k}+\phi_{i,j,k\pm 1}-6\phi_{i,j,k},
\label{Eqdifphi}
\end{equation}

\noindent
where $(i,j,k) =1,...,N_{\rm grid}$.
This leads to a large system of linear equations relating unknown
variables $\phi_{i,j,k}$ with known right-hand side of the discrete
form of the Poisson equation $3\Omega_0(\rho_{i,j,k}-1)/2a$. 
The system of equations is solved exactly by the FFT technique.

The density on the mesh $\rho_{i,j,k}$ is obtained from particle
positions using the Cloud-In-Cell method. In order to find the
``acceleration'' $\vc{g}=-\nabla\phi$ for each particle, the gravitational
potential is differentiated on the mesh:
\begin{equation}
g_x=-(\phi_{i+1,j,k}-\phi_{i-1,j,k})/2,\quad g_y=...,\quad g_z=...
\label{Eqgx}
\end{equation}
Then the acceleration is interpolated to the position of the particle
using a three-linear interpolation.
This scheme for the force interpolation (the same interpolation as in
the density assignment) is very important because it does not produce a
force acting on the particle itself. (Thus, an isolated point does not
produce a force at the position of the particle). While this might
sound like a natural condition for any realistic method, only two
methods -- PM and TREE -- do not have this self-force.  In P$^3$M the
effect is minimized. In the case of multigrid methods the self-force
cannot be excluded -- only minimized. Typically this is achieved by
placing extended buffer zones around regions with high resolution
(e.g. Kravtsov 1997). No precautions were made in AP$^3$M method,
which might result in spurious effects in regions were multi-level
grids are introduced.

Thus, the main scheme of the PM method consists of the following four
blocks repeated every time step:
\begin{itemize}
\item Find density on the mesh using the Cloud-In-Cell technique.
\item Solve the Poisson equation using two three-dimensional FFTs.
\item Advance velocities and coordinates of the particles.
\item Advance time and print results.
\end{itemize}

\section{\textbf{Format of data}}

In order to have the best possible resolution, most of the available
computer memory is allocated to the density/potential grid. The
Poisson solver is organized in such a way that only one large mesh is
needed.  Particle coordinates and velocities are kept on 
disk and are read into memory in large portions when necessary. This
reading/writing of particles results in a small overhead -- typically
5-10\% of the total cpu time. If this overhead is an issue, the code
can be easily adjusted to keep all particles in memory. This is
always the case for a parallel version of the code. Particles are
divided into ``species'' with constant mass of a particle for each
species.  Each species is kept in a separate file. Information which
describes the run (such as the number of particles, omegas, and current
time) is written in a separate header file.

Each file with particle data is a FORTRAN direct-access file with the number
of records equal to the number of particles in one direction $N_{\rm
row}$. Each record has coordinates and velocities for a ``page'' of
particles $N_{\rm page}=N_{\rm row}^2$. The ``page'' of particles is
read into a common block, which has the structure: $X(N_{\rm
page}),Y(N_{\rm page}),Z(N_{\rm page}), V_x(N_{\rm page}),V_y(N_{\rm
page}),V_z(N_{\rm page})$
The particle files and the header file are needed for
continuation of the run or for the data analysis.
The following diagram shows the structure and names of the data files:

\begin{tabular}{llll}
C3CRD.DAT& Header& & Text-of-Header,~$a$,~$a_{\rm init}$,
                        ~$\Delta a$, Step,$\ldots$\\ 
 & & & \\
C3crs0.DAT& Set 0& Page 1 & $x_1,x_2,\ldots x_{\rm Npage},$~$y_1, \ldots~
                           z_1, \ldots~ V_{x 1}, \ldots~ V_{z {\rm Npage}}$ \\
    &   & Page 2 & $x_1,x_2,\ldots x_{\rm Npage},$~$y_1, \ldots~
                    z_1, \ldots~ V_{x 1}, \ldots~ V_{z {\rm Npage}}$ \\
    &   & \hfil$\ldots$\hfil & \hfil$\ldots$\hfil \\
    &   & \hfil$\ldots$\hfil & \hfil$\ldots$\hfil \\
    &   & Page $N_{\rm row}$ & $x_1,x_2,\ldots x_{\rm Npage},$~$y_1, \ldots~
                           z_1,\ldots~ V_{x 1}, \ldots~ V_{z {\rm Npage}}$ \\
 & & & \\
C3crs1.DAT& Set 1& Page 1 & $x_1,x_2,\ldots x_{\rm Npage},$~$y_1, \ldots~
                           z_1, \ldots~ V_{x 1}, \ldots~ V_{z {\rm Npage}}$ \\
        &      & Page 2 & $x_1,x_2,\ldots x_{\rm Npage},$~$y_1, \ldots~
                           z_1, \ldots~ V_{x 1}, \ldots~ V_{z {\rm Npage}}$ \\
        &      & Page $N_{\rm row}$ & $x_1,x_2,\ldots x_{\rm Npage},$~$y_1, \ldots~
                           z_1, \ldots~ V_{x 1}, \ldots~ V_{z {\rm Npage}}$ \\
 & & & \\
$\ldots$&&&\\
$\underbrace{\hspace{1in}}$&\multicolumn{2}{l}{$\underbrace{\hspace{1.5in}}$}&$\underbrace{\hspace{3in}}$\\
File Name& \multicolumn{2}{c}{Description} &\hspace{1in}Content of the file\\
\multicolumn{3}{c}{Not a part of the file}\\
\end{tabular}

\noindent
Thus, the memory required to run the code is about $N_{\rm
grid}^3+6N_{\rm row}^2$ memory words or 64Mb$(N_{\rm grid}/256)^3$
+0.375Mb$(N_{\rm row}/128)^2$ if single precision arithmetic is
used. The total amount of disk space is \break 48Mb$(N_{\rm row}/128)^3$ per
each set of ``species''.

\section{\textbf{Initial conditions: CDM and CHDM models}}

We use the Zeldovich approximation to set initial conditions. The
approximation is valid in mildly nonlinear regime and is much 
superior to the linear approximation. We slightly rewrite the original
version of the approximation to incorporate cases (like CHDM) when the
growth rates $b(t)$ depend on the wavelength of the perturbation $|k|$. In the
Zeldovich approximation the comoving and the lagrangian coordinates are
related in the following way:

\begin{equation}
\vc{x} =\vc{q} -\alpha\sum_{\vc{k}}b_{|k|}(t)\vc{S}_{|k|}(\vc{q}),
\quad 
\vc{p} = -\alpha a^2\sum_{\vc{k}}b_{|k|}(t)
        \left( {\dot b_{|k|}\over b_{|k|}}\right)\vc{S}_{|k|}(\vc{q}),
\label{EqZeldtwo}
\end{equation}

\noindent
where the displacement vector $\vc{S}$ is related to the velocity
potential $\Phi$ and the power spectrum of fluctuations $P(|k|)$:

\begin{equation}
\vc{S}_{|k|}(\vc{q}) =\nabla_q\Phi_{|k|}(\vc{q}), \quad 
\Phi_{|k|}=\sum_{\vc{k}}a_{\vc{k}}\cos(\vc{k}\vc{q}) +
b_{\vc{k}}\sin(\vc{k}\vc{q}),
\label{EqZeldthree}
\end{equation}

\noindent
where $a$ and $b$ are gaussian random numbers with the mean zero and
dispersion $\sigma^2=P(k)/k^4$:

\begin{equation}
a_{\vc{k}}=\sqrt{P(|k|)}\cdot {Gauss(0,1)\over |k|^2}, \quad 
b_{\vc{k}}=\sqrt{P(|k|)}\cdot {Gauss(0,1)\over |k|^2}.
\label{EqZeldfour}
\end{equation}

The parameter $\alpha$, together with the power spectrum $P(k)$, define
the normalization of the fluctuations.

We estimate the power spectrum $P(k)$ for a wide range of cosmological
models using a Boltzman code (Holtzman 1989). As compared with
the original version of the code, the current version
allows for more accurate estimates at high wavenumbers. For each cosmological
model the numerical data points were fitted using the following fitting formula:

\begin{equation}
P(k) = {k^n\exp(P_1) \over (1 + P_2k^{1/2} +P_3k +P_4k^{3/2}+P_5k^{2})^{2P_6}}.
\label{EqFit}
\end{equation}

\noindent 
The coefficients $P_i$ are presented in the file \textbf{cdm.fit} for a
variety of models. The errors of
the fits are smaller than 5\% in the power spectrum. The top panel in
Figure 1 shows the errors of the fits for CDM models ($\Omega_0=1$) with
a Hubble constant $H=50$km/s/Mpc. Errors at a level of $\sim$ 2\% level at 
$k\sim 3h\Mpc^{-1}$ and at $k\sim 30h\Mpc^{-1}$ are due to small mismatch in
approximations used at high wavenumbers. The fits smooth out the jumps
and, thus, provide better approximations to the real power spectra at
those large wavenumbers. The waves around $k\sim 0.1h\Mpc^{-1}$ are due
to acoustic oscillations in baryons. They are larger for larger
$\Omega_b/\Omega_0$ ratios. For very small $\Omega_b/\Omega_0$ the
errors introduced by using the fits are extremely small. Thus, if one can 
neglect (or smooth out)
the acoustic oscillations, the maximum errors of our fits are expected
to be smaller than 1--2\% in the power. The comparison of some of our
power spectra with the results from COSMICS (Bertschinger 1996) support
our conclusion. We recommend the use of the fits whenever it is possible.

\begin{figure}
\epsfxsize=0.7\hsize
\caption{(Top) Errors of the fits eq.(\ref{EqFit}) for the CDM
models ($\Omega_0=1$) with a Hubble constant $H=50$km/s/Mpc. Errors
at the $\sim$ 2\% level at $k\sim 3h\Mpc^{-1}$ and at $k\sim 30h\Mpc^{-1}$ are
due to a small mismatch in approximations used at high wavenumbers. The
fits smooth out the jumps and, thus, provide better
approximations to the real power spectra at these large wavenumbers.
The waves around $k\sim 0.1h\Mpc^{-1}$ are due
to acoustic oscillations in baryons.
(Bottom) The differences between the power spectrum given by the BBKS
approximation and the power spectrum obtained from our fits. Triangles
show results obtained using COSMICS for $\Omega_b=0.05$}
\centering\leavevmode
\epsfbox{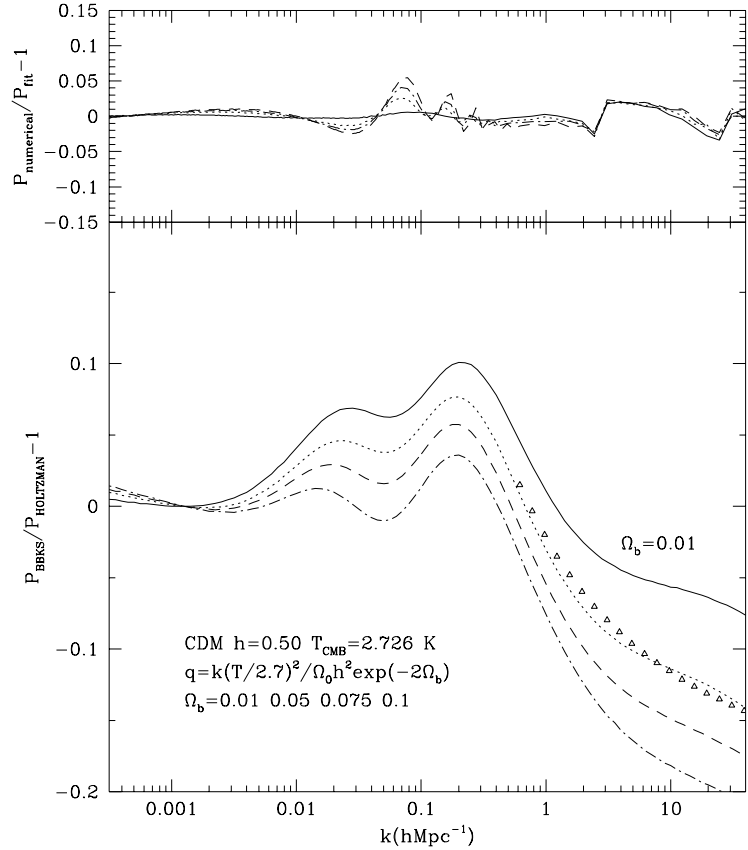}
\end{figure}
The power spectrum of cosmological models is often approximated using
a fitting formula given by Bardeen \etal (1986, BBKS):
\begin{equation}
P(k)=k^nT^2(k), \quad T(k)={\ln(1+2.34q)\over 2.34q}
     [1+3.89q+(16.1q)^2+(5.4q)^3+(6.71q)^4]^{-1/4},
\label{EqBBKS}
\end{equation}
\noindent 
where $q=k/(\Omega_0h^2\Mpc^{-1})$. Unfortunately, the
accuracy of this approximation is not great. Peacock \& Dodds (1994) modified
the fit using another relation between $q$ and $k$:

\begin{equation}
q=k/(\Omega_0h^2\exp(-2\Omega_b)\Mpc^{-1}).
\label{EqPDodds}
\end{equation}

This approximation was
criticized by Sugiyama (1995), who introduced a better scaling for
low-$\Omega_0$ cases:
\begin{equation}
 q={k(T_{\rm CMB}/2.7K)^2 \over
\Omega_0h^2\exp(-\Omega_b-\sqrt{h/0.5}\Omega_b/\Omega_0)\Mpc^{-1}}.
\label{EqSug}
\end{equation}

 \noindent 
These approximations have been frequently used in a number of publications
(e.g. Liddle \etal 1996). The bottom panel of Figure 1 shows the ratio
of the power spectrum given by this approximation to the power spectrum
obtained from our fits for several choices of baryon fraction. 
For comparison, we also present the error of
the eqs.(\ref{EqBBKS}-\ref{EqSug}) relative to the power spectrum obtained by
COSMICS for $\Omega_b =0.05$ (triangles), showing the good agreement of
our results with those of COSMICS.  
In all cases, there is a large decline (around
20\% in power) between a peak at $k=0.2h\Mpc^{-1}$ and small scales
$k\sim (10-30)h\Mpc^{-1}$. This decline was noticed by Hu \& Sugiyama
(1996), who studied the small-scale perturbations.  Note that if we
take $T_{\rm CMB}=2.70$K instead of 2.726K, than the peak of the error
at $k=0.2$ increases up to 15\%.  The error in the power is rather
small for small $k < 0.1h\Mpc^{-1}$ and for a realistic amount of
baryons $\Omega_b\sim 0.07$. One can easily miss it if instead of an
error of the power spectrum, one plots the transfer function in a
double logarithmic scale. But the error is very significant for
galactic-scale events. It can result in serious errors in the epoch of
galaxy formation or in the amount of gas in damped Ly-$\alpha$ clouds
at high redshifts.

Hu \& Sugiyama (1996) recommend changing the last parameter in the
BBKS fit from 6.71 to 6.07. We do not find that this correction gives
an accurate fit to our spectrum. We find that the following
approximation, which is a combination of a slightly modified BBKS fit
and the Hu \& Sugiyama (1996) scaling with the amount of baryons, provides
errors in the power spectrum smaller than 5\% for the range of
wavenumbers $k= (10^{-4} - 40)h\Mpc^{-1}$ and for $\Omega_b/\Omega_0<0.1$:

\begin{eqnarray}
P(k)  &=&k^nT^2(k), \nonumber\\
 T(k) &=&{\ln(1+2.34q)\over 2.34q}
     [1+13q+(10.5q)^2+(10.4q)^3+(6.51q)^4]^{-1/4},\nonumber\\
 q    &=&{k(T_{\rm CMB}/2.7K)^2\over
        \Omega_0h^2\alpha^{1/2}(1-\Omega_b/\Omega_0)^{0.60}},\qquad
 \alpha = a_1^{-\Omega_b/\Omega_0}a_2^{-(\Omega_b/\Omega_0)^3}\nonumber\\
 a_1 &=& (46.9\Omega_0h^2)^{0.670}[1+(32.1\Omega_0h^2)^{-0.532}],\quad
 a_2 = (12\Omega_0h^2)^{0.424}[1+(45\Omega_0h^2)^{-0.582}]
\label{EqUgly}
\end{eqnarray}

\noindent Figures 2 and 3 show errors of the fits for the CDM and for the
$\Lambda$CDM models.

\begin{figure}
\epsfxsize=0.8\hsize
\caption{Errors of the approximation eqs.(\ref{EqUgly}) for the
 CDM models with different Hubble constants and amount of baryons.}
\centering\leavevmode
\epsfbox{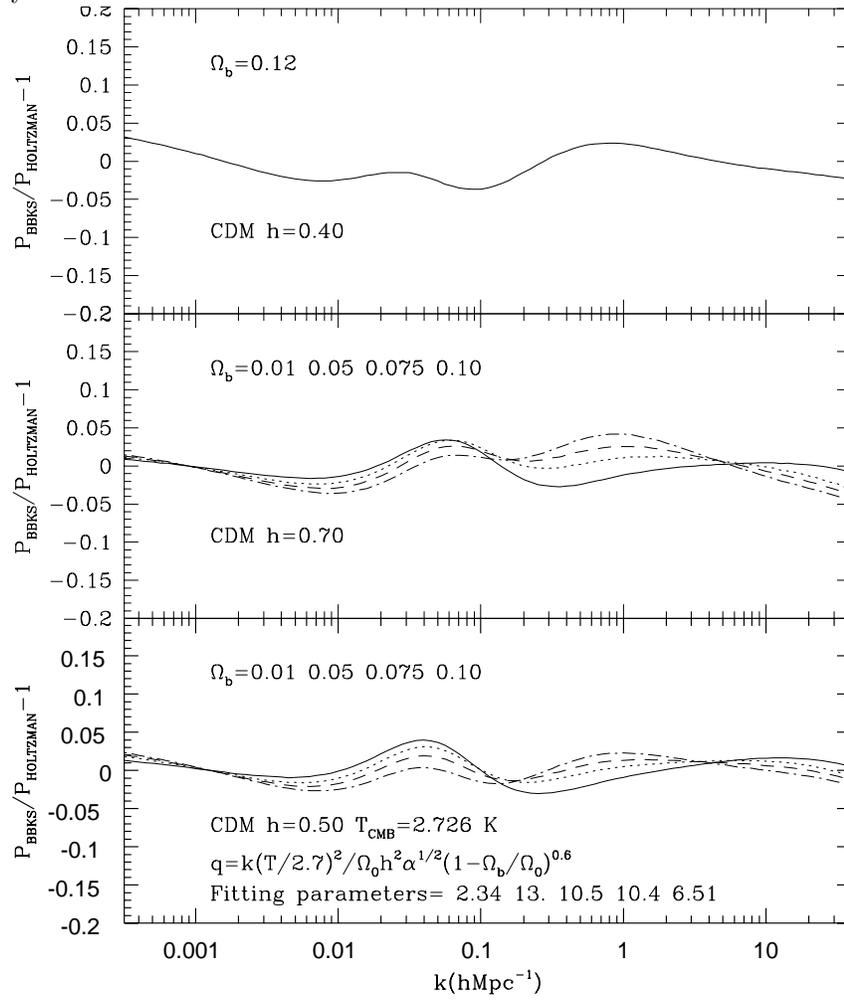}
\end{figure}

\begin{figure}
\epsfxsize=0.8\hsize
\caption{The same as Figure 2, but for the \char'3CDM models.  }
\centering\leavevmode
\epsfbox{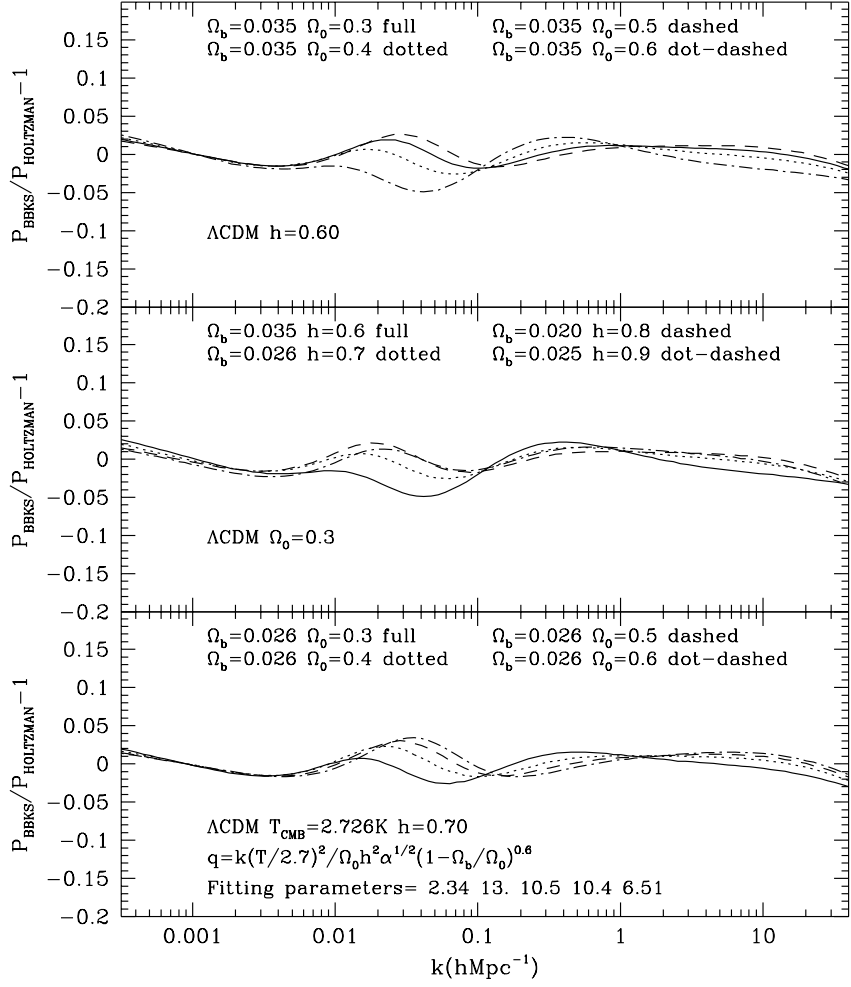}
\end{figure}

\section{\textbf{Finding Halos with Bound Density Maxima code}}

Finding halos in dense environments is a challenge. The most widely
used halo-finding algorithms -- the friends-of-friends (e.g., Efstathiou
\etal 1985) and the spherical overdensity algorithm (e.g., Lacey \&
Cole 1994; Klypin 1996) -- are not acceptable (Gelb \& Bertschinger
1994, Summers \etal 1995). The friends-of-friends (FOF) algorithm
merges together apparently distinct halos if the linking radius is too
large or misses some of the halos if the radius is small.  Adaptive FOF
(van Kampen 1995) seems to work better. But we find that it is
difficult to find an optimal scaling of the linking radius with the
density. We have developed a related algorithm (Klypin, Gottlober, Kravtsov
1997), which we call ``hierarchical friends-of-friends''. Because it
uses all linking radii, it does not have the problem that the adaptive FOF
algorithm has.  The algorithms, either adaptive or hierarchical, can
not work because they pick up many fake halos in very dense
environments. Klypin \etal (1997) supplement the hierarchical FOF algorithm
with an algorithm which checks if halos existed at previous moments.
The algorithm which finds halos as maxima of mass inside spheres of a
given overdensity works better than plain FOF, but no fixed
overdensity limit can find halos in both low and high density
environment.  The DENMAX algorithm (Gelb
\& Bertschinger 1994) and its offspring, SKID (Governato \etal 1997),
make significant progress -- they remove unbound particles, which is
important for halos in groups and clusters. Recently, Summers \etal
(1995) tried to perfect the idea of Couchman \& Carlberg (1992) to
trace the history of halo merging and to use it for halo
identification. Starting at an early epoch, Summers \etal identify
halos using the FOF algorithm with a linking radius corresponding to the
``virial overdensity'' of 200 and then trace particles belonging to
halos at later times. It appears that it is impossible to make a
working algorithm because halos interact too violently. A large fraction of
mass is tidally stripped from some halos and a large fraction of mass
is accreted.  Some of the problems that any halo finding algorithm
faces are not numerical. They exist in the real Universe. We select a few
typical difficult situations.

1. {\it A large galaxy with a small satellite.} Examples: LMC and the
Milky Way or the M51 system.  Assuming that the satellite is bound, do
we have to include the mass of the satellite in the mass of the large
galaxy? If we do, then we count the mass of the satellite twice: once when
we find the satellite and then when we find the large galaxy. This does
not seem reasonable. If we do not include the satellite, then the mass
of the large galaxy is underestimated. For example, the binding energy
of a particle at the distance of the satellite will be wrong. The
problem arises when we try to assign particles to different halos in
an effort to find masses of halos. This is very difficult to do for
particles moving between halos. Even if a particle at some moment has
negative energy relative to one of the halos, it is not guaranteed that
it belongs to the halo. The gravitational potential changes with time,
and the particle may end up falling onto another halo. This is not just
a precaution. This actually was found very often in real halos when we
compared contents of halos at different redshifts. Interacting halos
exchange mass and lose mass. We try to avoid the situation: instead of
assigning mass to halos, we find the maximum of the ``rotational
velocity'', $\sqrt{GM/R}$, which is observationally a more meaningful
quantity.

2. {\it A satellite of a large galaxy.} The previous situation is now
viewed from a different angle. How can we estimate the mass or the
rotational velocity of the satellite? The formal virial radius of the
satellite is large: the big galaxy is within the radius. The rotational
velocity may rise all the way to the center of the large galaxy. In
order to find the outer radius of the satellite, we analyze the density
profile. At small distances from the center of the satellite the
density steeply declines, but then it flattens out and may even
increase. This means that we reached the outer border of the
satellite. We use the radius at which the density starts to flatten out
as the first approximation for the radius of the halo. This
approximation can be improved by removing unbound particles and
checking the steepness of the density profile in the outer part.

3. {\it Tidal stripping.} Peripheral parts of galaxies, responsible for
extended flat rotation curves outside of clusters, are very likely 
tidally stripped and lost when the galaxies fall into a cluster. The
same happens with halos: a large fraction of halo mass may be lost due
to stripping in dense cluster environments. 
Thus, if an algorithm finds that 90\% of mass of a halo identified at
early epoch is lost, it does not mean that the halo was destroyed. This
is not a numerical effect and is not due to  ``lack of physics''. This
is a normal situation. What is left of the halo, given that it still has a large
enough mass and radius, is a ``galaxy''.

We have developed our halo-finding algorithm (Klypin \etal 1997) having
in mind all these problems.
The bound-density-maxima (BDM) algorithm stems from the DENMAX
algorithm (Gelb \& Bertschinger 1994). Just
as in DENMAX, the BDM algorithm first finds positions of the density
maxima on some scale and then removes unbound particles inside the halo
radius. However, the algorithm finds maxima and removes unbound
particles in a way different from DENMAX.  The algorithm can work
by itself or in conjunction with the hierarchical FOF. In the latter
case, it takes positions of halos from the hierarchical
friends-of-friends, and then removes unbound particles and finds
parameters of halos.

In order to find positions of halos we choose a radius $r_{sp}$ of a
sphere for which we find maxima of mass. This defines the {\it scale}
of objects we are looking for, but not exact radii or masses of
halos. The radius of a halo can be either larger or smaller than $r_{sp}$,
but distances between halos cannot be smaller than $r_{sp}$.  We place
a large number of spheres in the simulation box. The number of the
spheres is typically an order of magnitude or more larger than the
expected number of halos. For each sphere we find the mass inside the
sphere and the center of the mass. The center of the sphere is displaced to
the center of mass and the new mass and the center of mass is
found. The process is iterated until convergence. Depending on specific
parameters of the simulations, the number of iterations ranges from 10
to 100. This process finds local maxima of mass within sphere of radius
$r_{sp}$.

The efficiency of finding local maxima of mass depends on how the spheres
are chosen. In the present version of the code two algorithms were
implemented. (1) A small fraction of all particles are chosen as centers
of the spheres. The code asks you to enter $N_{seed}$, the ``Number of
particles for initial seeds''. Then it will select every
$N_{particles}/N_{seed}$ particle as a center of a sphere. (2)
Additional spheres will be placed in regions with relatively low
density. The whole simulation box is divided into a mesh of large
cells. The size of the cells is defined by the variable ``Cell'' in 
\textbf{PMhalos.f}. The ``Cell'' is typically equal to one or two PM cells.  If
a cell has many particles in it (``neighbors''), than some of them (not
more than 3) will be chosen as centers of spheres. The code asks you to
enter the minimum number of neighbors: ``Number of neighbors for a
seed''.

In some cases one would need to improve the location of the halo. An
example is if one is looking for groups of ``galaxies'' but also would
like to have the groups always centered on a galaxy.  The search radius
for the groups may be chosen to be, say 500~kpc. Additional iterations
with a smaller radius of the sphere will find the galaxy-size halo closest to
the center of mass of the group and place the center of the group at
the ``galaxy''. In the BDM code this option is realized in the
following way. The code asks you to enter the ``smaller radius for
final halos''. If this radius $r_{small}$ is not equal to the search
radius $r_{sp}$, the code will do additional iterations by gradually
changing the search radius from $r_{sp}$ to $r_{small}$. If $r_{small}
=r_{sp}$, no additional iterations are made.

Some of the density maxima will be found many times because in the
process of maximizing the mass some of spheres converge on the same local
maximum. Spheres which find the same maximum are called
``duplicates''.  We remove duplicates and keep only one halo for each
maximum.  Halos with too small number of particles (typically 5--10)
and halos with too low central overdensity are removed from the final
list. Parameters which control the removal are supplied by the user.

Once centers of potential halos are found, we start the procedure of
removing unbound particles and finding the structure of halos.  We
place concentric spherical shells around each center. For each shell we
find the mass of the dark matter particles, the mean velocity, the velocity
dispersion relative to the mean, and the maximum of the rotational
velocity $V_{\rm max}=\sqrt{GM(r)/r}\mid_{max}$. In order to
determine whether a particle is bound or not, we estimate the escape
velocity at the distance $r$ of the particle from the halo center: 

\begin{equation}
V^2_{\rm escape}(r) \approx (2.15\times V_{\rm max})^2 {\ln(1+2r/r_{max})
          \over (r/r_{max}) },
\end{equation}

\noindent 
where $r_{max}$ is the radius of the maximum of the
rotational velocity. This expression for the escape velocity is valid
for a halo with the Navarro-Frenk-White density profile.  If the velocity
of a particle is larger than the escape velocity, it is assumed to be
unbound. We estimate the maximum rotational velocity $V_{max}$ and
radius of the maximum $r_{max}$ using the density profile for the
halo. Because $V_{max}$ and $r_{max}$ must be found before the unbound
particles are removed and because the mean velocity is also found using
all particles (bound and unbound), the whole procedure cannot be done
in one step. We start by artificially increasing the value of the
escape velocity by a factor of three.  Only particles above the limit
are removed. We find a new density profile, new mean velocities, and new
$V_{max}$ and $r_{max}$. The escape velocity is again increased, but
this time by a smaller factor. The procedure is repeated 6 times. The
last iteration does not have any extra factors for the escape velocity:
all unbound particles are removed.  Examples of halos identified by the
code are presented in Klypin \etal (1997).

Finding a halo radius is straightforward for isolated halos: increase the
radius of sphere untill the overdensity inside the sphere is equal to
some limiting value provided by the user. For
halos inside groups or clusters (halos inside halos) the algorithm is
more complicated. It consists of three steps: (i) It starts with
finding the radius of given overdensity limit (as for an isolated
halo). (ii) Then, the algorithm checks how the mean overdensity inside
given radius changes with the radius. It starts going from the very
central shell outwards. If the mean overdensity stops declining, the
algorithm assigns the radius to the halo radius. (iii) Now the
algorithm goes from this radius inwards and checks the slope $n$ of the
overdensity profile: $(M(r)/\langle M(r)\rangle )\propto r^{-n}$. If the
slope is too shallow (mass increases too fast with the radius), the radius of
the halo is decreased because most of the mass at this radius does not
belong to the halo. The radius is decreased untill the slope is steep
enough. The limit for the slope is equal to $n=1$. This is a rather
mild slope: for an isothermal sphere one expects $n=2$; for the
Navarro-Frenk-White profile $n=1.7-2.5$.

The BDM code will ask you to enter many parameters. A typical dialog may look as
follows:

\bigskip
Enter Min. Center Overdensity for Halos   $=>$ 340.   \hfill (1)

Enter Overdensity Threshold for Halos  ~~   $=>$ 340. \hfill (2)

Enter Minimum halo mass in Msun/h   ~~~      $=>$ 5.e+9 \hfill (4)

Enter comoving search radius(Mpc/h) ~~~      $=>$ 0.050 \hfill (5)

Enter smaller radius(Mpc/h) of final halos$=>$ 0.030    \hfill (6)

Enter min.radius for halos(Mpc/h)  ~~~~~~~     $=>$ 0.030 \hfill (7)

Enter fraction of DM particles (1/4,1/2,1)$=>$ 1         \hfill (8) 

Enter rejection velocity limit (V/Vescape)$=>$ 1.0       \hfill (9)

Distance to check for Velocity duplicates $=>$ 0.300      \hfill (10) 

Define duplicates if  (v1-v2)/Vrms   $<$ ~~~~~~~~~  0.10    \hfill (11)

Enter Comoving Box size(Mpc/h)    ~~~~~~~~    $=>$ 20.     \hfill (12) 

\bigskip

In lines (1-2), enter the minimum overdensity for halos. If both values are
equal, the code will find halos with the overdensity above the limits
you provided. You may choose only those halos which have higher
central density, if you enter  a larger number in the first line.  Only
halos with mass larger than the value entered in the third line and
radius in the line 7 will be kept. For debugging of the code, you may
read only 0.25, or 0.5 of all particles (line 8).  If you would like to
remove unbound particles as described above, enter 1 in line 9. In
order to ignore the option (no removing of unbound particles) enter a
number larger than 5.  Lines 10 and 11 are used for two parameters
needed for additional screening of duplicates.  In case of very large
halos, when the density profile in the center of the halo is rather
flat, the iteration of spheres stops when the spheres are still far one
from the other. The spheres have found the same object: masses, radii,
velocities of the `` halos'' are very close, but their positions are
slightly different.  Because the fake halos have very close velocities,
they can be identified. The parameter in the line 10 defines the maximum
distance (in units of $\Mpch$) within which the code will look for the
duplicates. If the difference in velocities
$\sqrt{(v_{x1}-v_{x2})^2+(._y)+(._z)}$ is less than the parameter in
line (11) as measured in units of maximum of the rms velocities for
particles in the halos, only largest will be kept.  You may ignore the
option by entering two zeros.

\section{\textbf{How to compile and run the code}}

The code consists of several different FORTRAN programs:
\begin{itemize}
\item PM\_to\_ASCII.f
\item PMhalos.f
\item PMmain.f
\item PMmodelCHDM.f
\item PMmodels.f
\item PMpower.f
\item PMselect.f
\item PMstartCDM.f
\item PMstartCHDM.f
\end{itemize}

A Makefile is provided to allow easy compilation of all of the programs.
You should edit the Makefile and put in your preferred compilation flags
for:

\begin{itemize}
\item Optimization. On both our Sun machines running Solaris, and our HP
running HPUNIX, we use -O3.

\item The routine \textbf{PMmodels.f} should be compiled using double 
precision for all ``real'' variables and constants. On the Sun, this
can be accomplished using the -R8 flag, on the HP with -dbl.
\end{itemize}

The general scheme for running the code and analyzing results is as follows:
\begin{enumerate}
\item{ Set initial conditions using \textbf{PMmodels} and \textbf{PMstartCDM}.}

\item{ Run the PM code using \textbf{PMmain}.}

\item{ Analyze the results using \textbf{PMpower}, and \textbf{PMhalos}.
\textbf{PM\_to\_ASCII} will scale your results to ``normal'' units.}

\end{enumerate}

More details on each step are provided in subsequent sections.

\noindent Some important parameters and variables which are used in the codes,
for which you will either be prompted or may wish to change \textit{before
compilation}. \textbf{Please note the last item which may require you to
make a change in the code!}
\begin{itemize}
\item AEXPN = expansion parameter ($=1/(1+z)$). AEXP0 is the
expansion parameter at initial moment.
\item NROW = number of particles in one dimension ($=2^n$). The total
number of particles is equal to NROW$^3$. Particles are stored in
direct-access files. Each record of the files contains NROW$^2$ particles.
\item NGRID = size of the computational mesh ($=2^m$). The total number of
cells is NGRID$^3$. If you need to change either NROW or NGRID, the
only place where you need to make changes is the file \textbf{PMparameters.h}.
\item ASTEP = step in the expansion parameter $a$. Time integration is
done with a constant step in the expansion parameter: $da =$ASTEP.
\item ISTEP = current integration step
\item Nspecies = number of additional species of particles. This is
curretly used for the CHDM models. For plain CDM, ODM, or $\Lambda$CDM
models Nspecies = 0.
\item Om0,Oml0,Ocurv = densities of matter, cosmological constant,
and curvature at $z=0$.
\item hubble = the Hubble constant in units of $100$km/s/Mpc
\item NACCES = length of a record of direct-access files with
coordinates and velocities of particles (e.g., PMcrs0.DAT). For some
computers the length of the record is counted in bytes, for some it is
in machine words. If you get errors when running the code that
indicate problems with accessing the data files, multiply NACCES by 4
in file \textbf{PMauxiliary.f}, routine RDTAPE, and in files 
\textbf{ PMstart...f}. If the length of the data files is too long (it should be
$24\times NROW^3$ bytes), divide NACCES by 4. 
\end{itemize}

\section{\textbf{How to set initial conditions}}

Two programs need to be run in order to set initial conditions. If the
model you are interested in do not have massive neutrinos, 
the programs are \textbf{PMmodels} and \textbf{PMstartCDM}. For models
with massive neutrinos, the corresponding programs are
\textbf{PMmodelCHDM} and \textbf{PMstartCHDM}. Additionally, if you wish
to run the PM code for a model for which the initial power spectrum is
not included in the package, it is possible for you (with a bit more
work) to specify whatever power spectrum you want.

\subsection{Models without massive neutrinos}
A range of initial spectra which have been computed by our Boltzmann code
and fit with our fitting function (ea.\ref{EqFit}) are available. The
fitting parameters are tabulated in the file \textbf{cdm.fit}. You will
need to look in this file and determine the line number of the model which
you are interested in. The
first program (\textbf{PMmodels}) will ask to give you the following
parameters:

\begin{itemize}
\item Name of a file where it writes parameters
of the model and normalized power spectra.

\item $\sigma_8=$ rms density perturbation in a sphere of
radius $8h^{-1}$Mpc and the slope of the power spectrum at long 

\item Line number in the file \textbf{cdm.fit}

\item Size of the simulation box, number of particles in 1D,
and the redshift at which you would like to start your code (or to get
the power spectrum).

\item
The program will ask you if you need to have a file with all input
parameters needed to set initial conditions. If you need the file,
answer ``Yes'', and it will create file \textbf{InStart.dat}, which you
will give as input to the second code \textbf{PMstartCDM}. It will also
ask you few questions needed to produce the  file. If you answer
``No'', the code finishes its work by creating the file with the name
you gave it in the first line. The file has parameters of the model
(all $\Omega$'s, the Hubble constant, the age of the universe, growth
rate of density $\delta$, and $d(\ln{\delta})/d(\ln{a})$ at different
redshifts.  It also gives the bulk velocity of a sphere of radius
$50\Mpch$ and normalized power spectra of dark matter at redshift zero
($a=1$) and at the redshift you provided.

\item If you answered ``Yes'' for the previous question (you
needed to have an input file for \textbf{PMstartCDM}), you will be asked
to provide the following information:

\begin{itemize}
\item {\it A string of up to 45 characters (``header'')}. The header
is not used for simulations, but it is useful to label your
simulation. You can provide any information you want. This header will
stick to your run. All files with snapshots of your simulation and all
files with analysis of your simulation will have the header. Experience
of running many simulations shows that one never has enough information
describing details of a simulation done some time ago to identify that
simulation later. Use the header to identify your run.

\item {\it Step in the expansion parameter $da$.} This defines
 how many integration steps the code will do untill it runs to the
 redshift zero. If you would like to make $N$ steps and you start at
 redshift $z$, $da=(1-{1\over 1+z})/N$. The step $da$ should be
 (significantly) smaller than the initial expansion parameter $a_{init}
=-{1\over 1+z}$. 

\item {\it Seed for random numbers}. Use any integer number in
the range $1-(2^{31}-1)$ $(2^{31}\approx 2.1478\times 10^9)$.
\end{itemize}

\end{itemize}

Next you will run \textbf{PMstartCDM} which is the program that actually 
generates initial conditions for the PM code. It will generate two
files, which \textbf{PMmain} reads and updates: \textbf{PMcrd.DAT}
(information on the cosmological model and parameters of the run) and
\textbf{PMcrs0.DAT} (coordinates and velocities of particles). If you run
a CHDM simulation, you will have more data files with data for hot
neutrinos.  \textbf{PMstartCDM} will ask you to provide some parameters. If you
created file \textbf{InStart.dat} using the program \textbf{PMmodels}, simply
provide the file as input: $PMstartCDM < InStart.dat$. 

\subsection{Models with massive neutrinos}

In order to set initial conditions for a CHDM model, run \textbf{PMmodelCHDM}
and then \textbf{PMstartCHDM}. The codes will ask you similar questions as
for non-CHDM models. Because setting initial conditions for the CHDM
model is more complicated, we provide only one particular CHDM variant:
a model with two equal mass neutrinos with total contribution
$\Omega_{\nu}=0.20$, $h=0.5$. Initial conditions can be set only at
redshift $z=30$. 

\subsection{Initial conditions for arbitrary initial power spectra}

There are two ways of building initial conditions for models which are
not provided with the package. 
\begin{enumerate}

\item You may add a line to \textbf{cdm.fit}
with fitting parameters for your model and with parameters of approximation for
the power spectrum. This requires that your model is well fit by our
fitting formula (eq. \ref{EqFit}). The format of the \textbf{cdm.fit} file 
is described in the file. See also the routine \textbf{TRUNF(k)} for 
details of the approximation of the power spectrum of perturbations.  
\item You may change the functions TRUNF in \textbf{PMstartCDM.f} and Pk 
in \textbf{PMmodels.f} to return whatever initial spectra you desire.
\end{enumerate}

\section{\textbf{How to run the PM code}}

After you generate the initial conditions, you can start running
the code. Check if two files \textbf{PMcrd.DAT} and \textbf{PMcrs0.DAT}
were generated and are in your directory: \textbf{PMmain} will read the
files. It will also overwrite the files when it finishes. So, if you
need to have the original files, please make copies before you start
running \textbf{PMmain}. 

\section{\textbf{How to get coordinates and velocities of particles}}

To get the final coordinates and velocities, run \textbf{PM\_to\_ASCII}. 
This will convert the input files
\textbf{PMcrd.DAT} and \textbf{PMcrs0.DAT} into readable output files.
You will specify the output file name.

\section{\textbf{How to get power spectrum and density distribution}}

You can get the power spectrum and density distribution of your output
by running \textbf{PMpower}. This will read the raw output files from
\textbf{PMmain}, \textbf{PMcrd.DAT} and \textbf{PMcrs0.DAT}. The output
will go into a file called \textbf{Spectrum.DAT}.

\section{\textbf{How to find halos using the Bound-Density-Maxima code}}

\noindent Run \textbf{PMhalos}. The code will ask you many questions. It will
produce two files with the final results. ``Catalog.DAT'' contains
detailed information about all halos found by the code. After a rather
long preamble, data on the halos follow. Each halo has a ``header'',
which gives global parameters: coordinates, velocities, mass, and so on
for the halo (format is given in the preamble). ``Catshort.DAT'' has a
shorter preamble and has only a list of the halo headers.

\noindent {\it Tips:} The code may miss some halos if one chooses wrong 
parameters!

\begin{enumerate}
\item The number of spheres (seeds) should be very large:
  100,000 -- 150,000. If it is too small, the code
  misses some of small halos (but not the big ones).
\item The number of particles in each shell for the halo profile should be large:
  5-6 per shell. The code needs the density profile of a halo to find
  radius, escape velocity and so on.  It may get confused if the
  profile is too noisy.
\item Radius of the first bin for the profile should not
  be too small: not less than 1/2 of your force resolution.
  It should also contain few particles ($>2$).
\item Outer radius should be large. If density does not
  decline enough (flat profile), the code thinks
  that this halo is a fluke. If your radius is too
  small such that only the  central part of a potential
  halo fits in the radius, and the density gradient
  is not steep, the code will kill your halo.
\end{enumerate}

\section{\textbf{Examples}}

We make available two examples of runs of the PM code with data analysis
which you can use for comparison with your own results if that is desired.
The first
example has files for the test with $32^3$ particles and $128^3$
mesh. The second example is for $128^3$ particles and $256^3$ mesh.
Both tests were done for a $\Lambda$CDM model with $h=0.7$. 

These are available as two separate tar files with all results or you
can also grab individual result files. Since the output data
files for the second example are quite large, the tar file for
this case does not include these data files.
The full set of files from the $32^3$ example can be downloaded from 
{http://astro.nmsu.edu/ $\sim$aklypin/PM/TEST32x128.tar.gz}. The files from
the  $128^3$ example (without data files) can be downloaded from:
{http://astro.nmsu.edu/$\sim$aklypin/PM/TEST128x256.tar.gz}. 

Individual output files from the two cases can be downloaded from:
{http://astro.nmsu.edu/$\sim$aklypin/ PM/TEST32x128} and 
{http://astro.nmsu.edu/$\sim$aklypin/PM/TEST128x256}. We 
recommend looking at the output plots (*.ps.gz).

\clearpage
\section{\textbf{References}}

\ps  Appel A., {\it SIAM J. Sci. Stat. Comput.}, \textbf{6}
	(1985) 85.

\ps  Aninnos P., Norman M., Clarke D.A., \apj \textbf{436}
	(1994) 11.

\ps Bardeen, J.M., Bond, J.R, Kaiser, N., Szalay, A., 1986, \apj, 304,
15 (BBKS).

\ps  Barnes J. and Hut P., {\it Nature} \textbf{324} (1986) 446.

\ps  Bertschinger, E., \& Gelb, J.M. 1991, Comp. Phys., 5, 164

\ps  Bouchet F.R. and Hernquist L., \apjs \textbf{68 } (1988) 521. 

\ps  Couchman H.M.P., \apj  \textbf{368 } (1991) 23.

\ps  Couchman, H.P.M., \& Carlberg, R. 1992, \apj, 389, 453

\ps  Efstathiou G., Davis M., Frenk C.S., and  White S.D.M.,
	\apjs \textbf{57} (1985) 241.

\ps  Gelb J., Bertschinger, E., 1994, \apj, 436, 467.  

\ps  Governato, F., Moore, B.,  Cen, R.,  Stadel, J., Lake, G., 
 Quinn, T., 1997, New Astronomy, 2, 91.

\ps  Gross, M., {\it Ph.D. Thesis,} UCSC  (1997)

\ps  Hernquist L., \apjs  \textbf{64 } (1987) 715.

\ps  Hernquist L., Bouchet F.R., and Suto Y., \apjs \textbf{75 }
	(1991) 231. 

\ps  Hockney R.W. and Eastwood J.W., {\it Numerical
        simulations using particles} (New York: McGraw-Hill) 1981.

\ps  Hu, W., Sugiyama, N., 1996, \apj, 471, 542.

\ps  Kates, R., Kotok, E., Klypin, A, 1991, Astron. \& Astrophys., 243, 295.

\ps  Klypin A., Holtzman J., Primack J., and Regos E., \apj
	\textbf{416} (1993) 1. 

\ps  Klypin, A. 1996, in ``Dark Matter in the Universe'', p. 419,
 eds. S. Bonometto, J.Primack, A. Provenzale, IOS Press, Amsterdam,
 Oxford, Tokyo, Washington DC

\ps Klypin, A, Nolthenius, R., Primack, J., 1997, \apj, 474, 533.

\ps Klypin, A., Gottlober, S., Kravtsov, A., Khokhlov, A., 1997, \apj,
submitted.

\ps Kravtsov, A.,  Klypin, A.,  \& Khokhlov, A. 1997,
   Astrophys. J. Suppl., 111, 73

\ps Lacey, C., Cole, S. 1994, \mnras, 271, 676

\ps Liddle, A.R., Lyth, D., Roberts, D., Viana, P.T.P., 1996, \mnras,
      278, 644. 

\ps  Melott A.L., {\it Phys. Rev. Letters} \textbf{56} (1986) 1992.

\ps  Peacock, J., Dodds, S.J., 1994, \mnras, 267, 1020. 

\ps  Suisalu, I., Saar, E., 1995, \mnras, 274, 287.

\ps  Sellwood J.A., {\it Ann. Rev. Astron. Astrophys.}
	\textbf{25} (1987) 151.

\ps Sugiyama, N., 1995, \apj, 471, 542.

\ps Summers, F.J., Davis, M., \& Evrard, A. 1995, \apj, 454, 1

\ps van Kampen, E. 1995, \mnras, 273, 295

\clearpage

\end{document}